\documentclass{iaus}
\usepackage{graphics}

  \checkfont{eurm10}
  \iffontfound
    \IfFileExists{upmath.sty}
      {\typeout{^^JFound AMS Euler Roman fonts on the system,
                   using the 'upmath' package.^^J}%
       \usepackage{upmath}}
      {\typeout{^^JFound AMS Euler Roman fonts on the system, but you
                   dont seem to have the}%
       \typeout{'upmath' package installed. iaus.cls can take advantage
                 of these fonts,^^Jif you use 'upmath' package.^^J}%
      }
  \else
  \fi

  \checkfont{msam10}
  \iffontfound
    \IfFileExists{amssymb.sty}
      {\typeout{^^JFound AMS Symbol fonts on the system, using the
                'amssymb' package.^^J}%
       \usepackage{amssymb}%

      }{}
  \fi

  \IfFileExists{amsbsy.sty}
    {\typeout{^^JFound the 'amsbsy' package on the system, using it.^^J}%
     \usepackage{amsbsy}}
    {}

\newsavebox{\astrutbox}
\sbox{\astrutbox}{\rule[-5pt]{0pt}{20pt}}

\title[Outskirts of Galaxy Clusters: intense life in the suburbs]
{Clusters outskirts at X-ray wavelengths: current status and future prospects}

\author[S. Molendi ]%
{Silvano Molendi }

\affiliation{Istituto di Astrofisica e Fisica Cosmica, Sezione di Milano,
Via Bassini 15, I-20133, Milano Italy, email: silvano@mi.iasf.cnr.it\\}

\pubyear{2004}
\volume{195}
\pagerange{1--8}
\date{?? and in revised form ??}
\jname{Outskirts of Galaxy Clusters: intense life in the suburbs}
\editors{A. Diaferio, ed.}
\begin{document}

\maketitle

\begin{abstract}

A solid observational characterization of cluster regions around
the virial radius would allow us to improve considerably our
understanding of the physics of galaxy clusters as a whole:
sadly current and planned experiments will not allow us to study
these regions. Unbeknownst to most but not all, the development
of an experiment sensitive to cluster outer regions could be
achieved with currently available technology, with no need
for breakthroughs.  Amongst the major factors that
will decide  whether and when such an experiment will be flown are
the awareness of the cluster community of the importance of such a
mission and its determination in supporting it.

\end{abstract}

\firstsection 

\section{Why study cluster outskirts?}

In the hierarchical Universe clusters form at the intersection of
cosmic filaments through accretion (see for example the picture
on the the conference poster taken from Borgani et al. 2004).
Shocks are expected to occur where the free-falling gas collides
with the ICM (i.e. Tozzi et al. 2000). These shocks convert the
bulk of the kinetic energy of the free-falling gas into thermal
energy, a fraction of the kinetic energy may be retained by the
shocked gas (i.e. Schuecker et al. 2004), playing a non trivial role 
in shaping cluster. 
The boundary between cluster and free falling gas is likely
not in a spherical form, as accretion occurs mainly from a few
preferential directions. The thermodynamic status of the
pre-shocked gas will likely determine the strength of the shock;
if, for example, the gas has been pre-shocked in filaments, as
envisaged in the pre-heating scenario, the shock will be weaker
than if it has not been pre-shocked (i.e. Tozzi et al. 2000).

Relatively massive in-falling substructures, such as groups, are
likely to retain at least part of their structure all the way down
to the cluster central regions (i.e. A3667, Vikhlinin et al. 2001). 
Shocks generated by these structures will
likely leave in their wakes turbulence which will eventually be
dissipated at smaller scales (i.e. Sunyaev these proceedings, 
Schuecker et al. 2004).
Less massive structures such as galaxies will suffer severe
modifications already in the cluster outskirts.  The compression
associated to the shock will amplify the strength of the magnetic
fields which are frozen in the high conductivity plasma,
particles will likely be accelerated at the site of the shock.

Ram pressure stripping of the metal rich gas from the in-falling
galaxies by the ICM (i.e. Treu, Moore these proceedings), will
enrich the ICM in heavy elements. Direct accretion of pristine
gas, never processed in galaxies, will have the opposite effect
of diluting the metal content of the ICM.

These simple considerations illustrate how important the regions
around the virial radius, the cluster outskirts, are. The rather
obvious reason for the occurrence of the wide variety of phenomena
sketched above is that these are the regions where virialization
occurs. Viewing the issue from an observer's point
of view we may  say that as long as we do not have a solid
observational characterization of cluster outer regions our
understanding of clusters  as a whole will necessarily be a
limited one.

\section {How much do available observations tell us?}

Let us then see how much of this extremely important
observational work we can carry out with currently available data.
Let us consider the three principle X-ray observables: the surface
brightness, the temperature and the metal abundance.

Measurements of  the surface brightness extend out to about 0.6
r$_{180}$ (Mohr et al. 1999, Ettori \& Fabian 1999, Ettori et al. 2004). 
The detector that has
been most successful in these measurements, at low redshifts, i.e.
z$<0.2$, is not part of the payload of the latest generation of
X-ray satellites, indeed it the ROSAT PSPC, which was launched 14
year ago.

ASCA (Markevitch et al. 1998), BeppoSAX (De Grandi \& Molendi 2002) 
and XMM-Newton (Pratt \& Arnaud 2002, Zhang et al. 2004) 
allow us to extend temperature measurements
out to about 0.5-0.6 r$_{180}$. Is is rather interesting that
XMM-Newton, with an effective area more than ten times larger than
ASCA or BeppoSAX is not capable of extending temperature
measurements beyond regions already explored by the latter
satellites. It is also a matter of some concern that the first
comparisons between the XMM-Newton and BeppoSAX temperature
points to a non-negligible disagreement (i.e. De Grandi \& Molendi
these proceedings).

Finally, BeppoSAX, ASCA  and XMM-Newton measurements of the metal
abundance extend out to about 0.4 r$_{180}$, i.e. De Grandi \&
Molendi (2001). This is indeed the most difficult observable to
measure from low surface regions, at least with the current CCD
resolution of a few percent.

Thus we are forced to conclude that current data does not go out
to the virial radius, as a matter of fact it does not even come
close to it. The answer to this sections  question is that 
\underline{we know nothing from direct X-ray observations of the  extremely
important region} \underline{around the virial radius.}

\section {How difficult would it be to design experiments capable of
investigating these regions?}

Having learned that current experiments are not sensitive to
cluster outer regions we ask ourselves how difficult it would be
to design experiments capable of investigating these regions.
The answer to this question is that it would not be particularly
difficult, no major technological breakthrough is required.
Indeed the answer lies in keeping the background under control.
Let us see how experiments could be optimized to do that.

\subsection{ NXB $<$ CXB }

The first thing to do is to reduce the non X-ray background
(NXB) to values smaller than the Cosmic X-ray background (CXB) in
the whole energy band covered by our experiment. Given the greater
difficulty in focusing higher energy X-rays and the flatness of the
NXB spectrum this typically means
having NXB $<$ CXB at the highest energies. It is worth remarking
that none of the imaging experiments sensitive in the medium
energy band (2-10 keV), flown so far, operate in the NXB $<$ CXB
regime. Neither the first generation experiments (BeppoSAX, ASCA),
nor the second generation experiments (Chandra, XMM-Newton).
As a matter of fact, the CXB/NXB ratio for the latter is worse than 
for the former and this is why XMM-Newton temperature 
profiles  do not extend further out that ASCA or Beppo-SAX temperature 
profiles.

Achieving the NXB $<$ CXB requirement would allow us to go about a factor
10 fainter in surface brightness. This is not really impossible,
in the case of XMM-Newton for example, simply placing the
satellite in an equatorial Low Earth Orbit would do more than half
of the job. The rest, a factor of about 2, could probably be
achieved by working on the shielding of the detectors.

\subsection{ Observing Strategy }

The other thing to do is to improve the observing strategy.
Typically temperature measurements can be made out to radii where
the source counts, cts$_{\rm sou }$, are equal to the background
counts, cts$_{\rm bkg }$. Measurements at larger radii cannot be
made because the background is only poorly known. In other words
we need to improve measurements in the cts$_{\rm sou } < $
cts$_{\rm bkg }$. This is where differential techniques can be
very effective. The most striking example comes from radio
wavelengths, indeed WMAP detects CMB anisotropies of 1 part in
$10^6$! In the X-ray band the most successful  experiment is 
the BeppoSAX PDS. This
experiment makes use of 2 rows of rocking collimator one is
trained on the source, the other on the background: the
collimators are switched every $\sim$  100s to allow the
detectors, viewed by both collimators, to observe source and
background fields.

\begin{figure}
\centering
\rotatebox{-90}{\resizebox{!}{14cm}{%
   \includegraphics{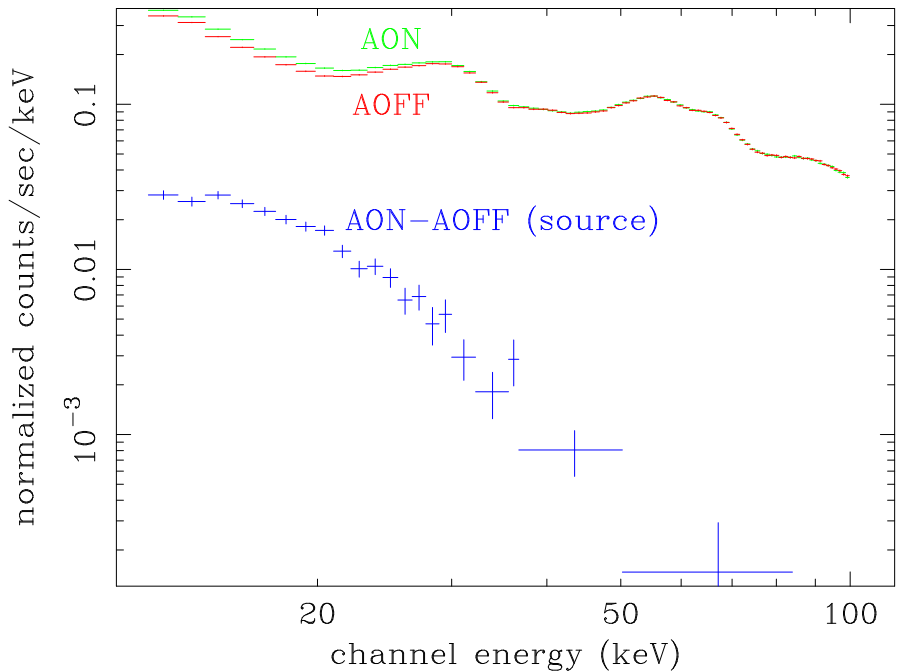}}}
  \caption{
ON-spectrum, OFF-spectrum and source spectrum
(difference between ON and OFF ) for collimator row A during a 
PDS pointing.
}
\end{figure}

 In Fig. 1 we show  the ON-spectrum, the
OFF-spectrum and the  source spectrum (difference between ON and
OFF ) for collimator row A during a PDS pointing. Extensive
analysis (i.e. Rossetti 2003, Rossetti \& Molendi 2004) have shown
that the source spectrum measurement is reliable down to where
source counts are about 1\% of the background counts. The
strategy adopted by the PDS could easily be implemented in future
imaging missions, simply by substituting collimators with telescopes.

Combining the two strategies (i.e. reducing the NXB intensity
below the CXB intensity and using differential measurements to
keep the background under control) we could easily go down by a
factor of 100 in surface brightness reaching the required
sensitivity to map outer regions of galaxy clusters.

\section {What about already planned missions?}

What about already planned X-ray missions, will they do better
than current ones in investigating cluster outskirts? Probably
not, missions such as ASTRO-E2, Con-X and XEUS are not optimized for
the study of low surface brightness emission. The general
approach is to achieve a high sensitivity through a large
effective area, paying relatively little attention to background
issues. This works well for point-sources where the background can
be kept under control through the size of the PSF, but is of
limited use for extended sources, where the instrumental
background intensity per unit solid angle increases with the
telescope focal length.

Of course we are not trying to imply that planned missions will
be useless for galaxy clusters.  They will certainly allow us to
study important cluster properties, as an example consider the
measurements of gas motions and turbulence that will be possible
with ASTRO-E2  (i.e. Sunyaev these proceedings). However these
studies will be limited to the brighter regions of galaxy
clusters. In a sense our investigations will be restricted to the
tip of the proverbial iceberg.

\section {Summary}
Observational characterization of outer regions is a fundamental
step in understanding clusters. Unfortunately current and planned
experiments will not allow us to study these regions. Fortunately
no major technological breakthroughs are required to design an
experiment sensitive to cluster outskirts. Whether and when such
an experiment will be designed and flown will depend upon many
factors. One of the most important will be the awareness of the
cluster community of the need for such a mission and its 
determination in supporting it.


\end{document}